# Ultra-high resolution multimodal MRI densely labelled holistic structural brain atlas

*José V. Manjón[1], Sergio Morell-Ortega[1], Marina Ruiz-Perez[1], Boris Mansencal[2], Edern Le Bot[2], Marien Gadea[3], Enrique Lanuza[4], Gwenaelle Catheline[5], Thomas Tourdias[6], Vincent Planche[7], Remi Giraud[8], Denis Rivière[9], Jean-Francois Mangin[9], Nicole Labra-Avila[9], Roberto Vivo-Hernando[10], Gregorio Rubio[11], Fernando Aparici[12], Maria de la Iglesia-Vaya[13,14] and Pierrick Coupé[2]*

[1] Instituto de Aplicaciones de las Tecnologías de la Información y de las Comunicaciones Avanzadas (ITACA), Universitat Politècnica de València, Camino de Vera s/n, 46022, Valencia, Spain

[2] CNRS, Univ. Bordeaux, Bordeaux INP, LABRI, UMR5800, in2brain, F-33400 Talence, France

[3] Department of Psychobiology, Faculty of Psychology, Universitat de Valencia, Valencia, Spain

[4] University Valencia, Department of Cell Biology, Burjassot, 46100, Valencia, Spain

[5] University Bordeaux, CNRS, EPHE, PSL, INCIA, UMR 5283, F-33000, Bordeaux, France

[6] Service de Neuroimagerie diagnostique et thérapeutique, CHU de Bordeaux, F-33000 Bordeaux, France

[7] Institut des Maladies Neurodégénératives, Univ. Bordeaux, CNRS, UMR 5293, F-33000 Bordeaux, France

[8] Univ. Bordeaux, CNRS, Bordeaux INP, IMS, UMR 5218, F-33400 Talence, France

[9] NeuroSpin, BAOBAB lab, CEA Saclay, Gif-sur-Yvette, France

[10] Instituto de Automática e Informática Industrial, Universitat Politècnica de València, Camino de Vera s/n, 46022, Valencia, Spain.

[11] Departamento de matemática aplicada, Universitat Politècnica de València, Camino de Vera s/n, 46022 Valencia, Spain.

[12] Área de Imagen Médica. Hospital Universitario y Politécnico La Fe. Valencia, Spain

[13] Unidad Mixta de Imagen Biomédica FISABIO-CIPF. Fundación para el Fomento de la Investigación Sanitario y Biomédica de la Comunidad Valenciana - Valencia, Spain.

[14] CIBERSAM, ISC III. Av. Blasco Ibáñez 15, 46010 - València, Spain

**\*Corresponding author:** José V. Manjón. Instituto de Aplicaciones de las Tecnologías de la Información y de las Comunicaciones Avanzadas (ITACA), Universidad Politécnica de Valencia, Camino de Vera s/n, 46022 Valencia, Spain.

Tel.: (+34) 96 387 70 00 Ext. 75275 Fax: (+34) 96 387 90 09.

E-mail address: jmanjon@fis.upv.es (José V. Manjón)

**Abstract**

In this paper, we introduce a novel structural holistic Atlas (holiAtlas) of the human brain anatomy based on multimodal and high-resolution MRI that covers several anatomical levels from the organ to the substructure level, using a new densely labelled protocol generated from the fusion of multiple local protocols at different scales. This atlas was constructed by averaging images and segmentations of 75 healthy subjects from the Human Connectome Project database. Specifically, MR images of T1, T2 and WMn (White Matter nulled) contrasts at 0.125 mm$^3$ resolution were selected for this project. The images of these 75 subjects were nonlinearly registered and averaged using symmetric group-wise normalisation to construct the atlas. At the finest level, the proposed atlas has 350 different labels derived from 7 distinct delineation protocols. These labels were grouped at multiple scales, offering a coherent and consistent holistic representation of the brain across different levels of detail. This multiscale and multimodal atlas can be used to develop new ultra-high-resolution segmentation methods, potentially improving the early detection of neurological disorders. We make it publicly available to the scientific community.

# 1. Introduction

The field of neuroscientific research has been revolutionized by the development of advanced imaging techniques, with Magnetic Resonance Imaging (MRI) standing at the forefront. MRI provides a non-invasive and high-resolution approach to investigate the details of the human brain anatomy. In particular, the development and utilization of MRI-based brain atlases have emerged as invaluable tools that play an important role to standardize image resolutions, orientations and label definitions, and to provide a common ground for brain research worldwide which has definitely helped in understanding the complex architecture of the brain.

A brain atlas serves as a reference framework that maps and delineates various anatomical and functional regions within the brain. It provides a standardized coordinate system, allowing researchers to precisely locate regions of interest at the individual level and also compare them across individuals or populations. The importance of MRI-based brain atlases lies in their multifaceted utility across diverse domains of neuroscientific investigation.

MRI-based brain atlases enable detailed structural mapping of the brain, allowing researchers to visualize and study the morphology of different regions. This aids in the identification of anatomical landmarks and exploring the spatial relationships between brain structures. In the clinical realm, MRI-based brain atlases are indispensable tools for accurate diagnosis and treatment planning. Neurosurgeons rely on these atlases to navigate through the brain during surgical procedures, ensuring precision in targeting specific areas while minimizing damage to surrounding healthy tissue. Brain atlases also serve as a common reference for population-based studies (Evans et al., 1993), allowing researchers to compare and contrast brain structures across diverse demographic groups to better understand the natural history of neurological diseases (Coupe et al., 2023; Planche et al., 2023;2024). They facilitate the integration of data from multiple studies, fostering collaborative efforts and meta-analyses to derive more robust conclusions.

Several brain atlases have been developed over the years, each serving distinct purposes in neuroscientific research, clinical applications, and medical practice. The Talairach and Tournoux Atlas (Talairach and Tournoux, 1988), originally developed for stereotactic neurosurgery, introduced a three-dimensional grid system based on post-mortem anatomical landmarks, serving as an early milestone for spatial normalization. However, modern neuroimaging requires more sophisticated and standardized approaches. The Montreal Neurological Institute (MNI) Atlas (MNI305) (Evans et al., 1993) addressed this need by offering a standardized coordinate system, facilitating meta-analyses and cross-study comparisons (in 2001 an improved version of this atlas named MNI152 (Mazziotta et al., 2001; Fonov et al., 2009; 2011) was proposed and is currently one of the most used ones). Hammer et al. (2003) proposed an atlas with 49 labelled structures. Shattuck et al. (2007) also proposed a probabilistic atlas of human cortical structures. Later, with a larger number of structures (n=90), the AAL (Automated Anatomical Labeling) Atlas (Tzourio-Mazoyer et al., 2002; Rolls et al., 2015; 2020) was proposed and has been widely used in many functional neuroimaging studies, providing automated labelling of brain regions for voxel-based analyses. More recently, the Brainnetome Atlas (Fan et al., 2016), was designed to help in the analysis of brain networks and connectivity, providing insights into functional and structural connections between different brain regions. From a more structural perspective, the Desikan-Killiany Atlas (Desikan et al., 2006) and the MINDboggle Atlas (Klein et al., 2017) have also been commonly used in structural MRI studies.

There are also brain atlases derived from histology that have exceptional detail. The Allen Human Brain Atlas (Hawrylyczvet al., 2012) (https://atlas.brain-map.org/) is focused on gene expression patterns in the human brain. The Allen Atlas is crucial for understanding the molecular organization of different brain regions and includes detailed maps of gene expression across the entire human brain, allowing researchers to explore the genetic basis of brain function. BigBrain (Amunts et al., 2013) is a model of a human brain at a cellular resolution of nearly 20 micrometers, based on the reconstruction of 7404 histological sections of a single brain. An updated probabilistic version, named Julich-Brain (Amunts et al., 2020) was recently proposed. However, these atlases are not MRI based and their use with clinical-quality MRI data can be challenging.

All these atlases, among others, serve as essential tools in neuroscientific research, providing a standardized framework for the interpretation and comparison of brain imaging data. Their diverse features meet several research needs, from structural and functional mapping to gene expression patterns and connectivity analysis. A compressive list of many of them can be found here (https://fsl.fmrib.ox.ac.uk/fsl/fslwiki/Atlases).

Currently, the MRI-based atlases have a maximum resolution of 1 mm$^3$ and typically use the T1w image modality due to its high anatomical contrast. However, higher image resolutions are starting to be available (Lusebrink et al., 2021; Stucht et al., 2015) thanks to either the use of new ultra-fast MR acquisitions (many of them powered nowadays by the use of artificial intelligence (Singh et al., 2023)) or by the use of post-acquisition super-resolution techniques (Zhanxiong et al, 2023; Grover et al., 2024). Initiatives to produce higher-resolution atlases are currently in development (Schira et al., 2023; Casamitjana et al., 2024). Also, multimodal approaches have been explored (Glasser et al., 2016).

In this paper, we propose a new structural holistic MRI-based ultra-high resolution multimodal densely labelled atlas. This atlas is based on ultra-high resolution *in vivo* MRI and has been labelled using the fusion of currently available tools for brain parcellation. The improved resolution of the atlas (0.125 mm$^3$ vs typical 1 mm$^3$), its multimodal nature and its dense and holistic labelling will facilitate the measurement of more subtle anatomical patterns and hopefully will contribute to earlier diagnostics and analyses. In the next sections, the atlas construction details and the resulting atlas are described.

## 2. Material and Methods

To construct the aforementioned structural atlas, we used images from a public dataset and a private dataset, and segmentations from existing tools as a starting point. In this section, the details of this process are summarized.

### 2.1. Dataset description

We used MR images from 2 datasets to construct the atlas. On one hand, we randomly selected 75 T1w and T2w images from healthy subjects of the Human Connectome project (HCP), specifically, the HCP1200 dataset (https://www.humanconnectome.org/study/hcp-young-adult/document/1200-subjects-data-release). Those images were taken in a 3T MR scanner from healthy subjects (41 females and 34 males) with ages between 22 and 35 years. High-resolution T1w and T2w images had a matrix size of 260x311x260 voxels and a voxel size of 0.7x0.7x0.7 mm$^3$. On the other hand, we used also 55 young healthy voluntary subjects from a private dataset acquired in a 3T scanner (Vantage Galan 3T/ZGO; Canon Medical Systems) in Bordeaux Hospital as a part of the DeepMultiBrain research project. In this dataset, each subject had T1w, T2w and WMn (White Matter nulled) images. In this dataset, T1w and T2w images had a matrix size of 256x376x368 voxels and a voxel size of 0.6x0.6x0.6 mm$^3$. WMn images had a matrix size of 448x548x400 voxels and a voxel size of 0.4x0.4x0.4 mm$^3$. WMn images have an excellent contrast for deep gray matter structures, especially useful for thalamic nuclei segmentation. This dataset was not used in the atlas construction but to generate synthetic WMn not available in the HCP dataset.

## 2.2. Image preprocessing

All the selected T1w and T2w images from HCP1200 dataset underwent a preprocessing stage to place them in a standard intensity and geometric space. This phase consists of several steps. First, both images were denoised using the Spatially Adaptive Non-local means (SANLM) filter (Manjón et al., 2010) and later inhomogeneity corrected using the N4 bias correction method (Tustison et al., 2010). The filtered images were then affine-registered to the MNI152 space at 0.125 mm$^3$ resolution (voxel size of 0.5x0.5x0.5 mm$^3$) using ANTs software (Avants et al., 2011). The resulting images have a standard matrix size of 362x434x362 voxels. Finally, the images were intensity normalized using the TMS method (Manjón et al., 2008).

For the DeepMultiBrain Bordeaux dataset, the same preprocessing was applied with the exception that rigid transformation from native T2w and WMn images to the native T1w was first estimated to later concatenate it with the affine transform of T1w image to map all the images to the same MNI152 space (HCP T1w and T2w images were already registered).

As we wanted to use public data in the creation of the atlas, we decided to synthetize the WMn images using HCP1200 data instead of using the DeepMultiBrain Bordeaux dataset. Currently, there are modern image synthesis techniques that make it possible to synthesize non-acquired modalities from others (Dar et al., 2019; Manjón et al., 2021). To generate WMn-like images in the HCP1200 dataset, a multimodal variant of a full-volume synthesis method (Manjón et al., 2021) was used. We trained this variant using the DeepMultiBrain Bordeaux dataset where the input images (T1w and T2w) were used to generate a WMn image. Once the network was trained, it was applied to the HCP dataset to generate the WMn images. Since this network works at 1 mm$^3$ resolution, it was applied 8 times to a volume-to-channel decomposition that transforms the input volume of 362x548x362 voxels into 8 volumes of 181x217x181 voxels using striding with step 2 at each dimension as done in method DeepICE (Manjón et al., 2020a). After synthetizing the 8 WMn volumes the process is reversed to obtain the final 362x548x362 voxel WMn volume. Figure 1 shows an example of the T1w, T2w and synthetized WMn images. The quality of the synthetized WMn images was validated by our experts.

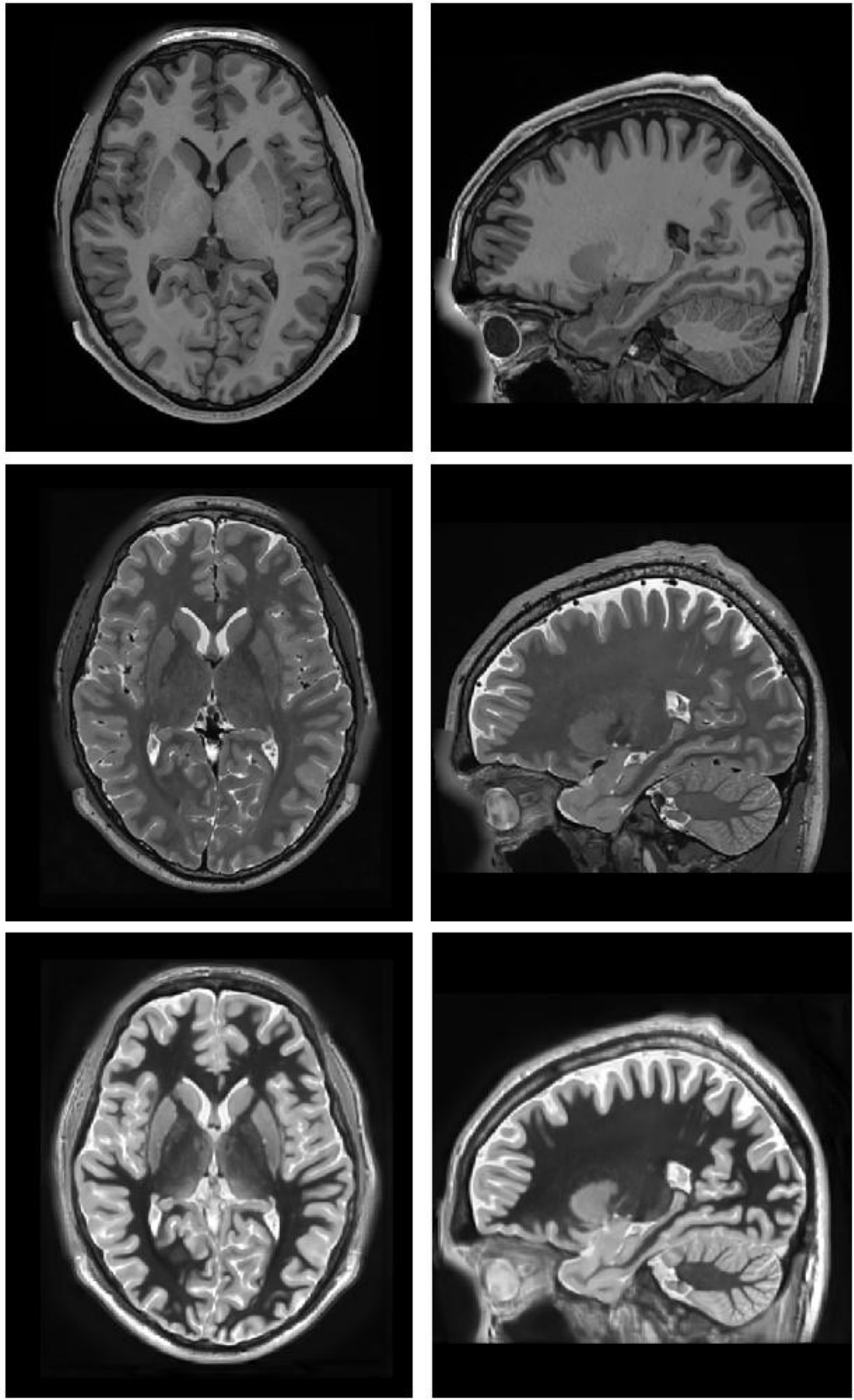

*Figure 1. Example of a preprocessed T1w, T2w and the synthetic WMn images at 0.125 mm³ resolution MNI152 space.*

**2.3. Software packages**

To generate a densely labelled atlas, we decided to merge several available protocols from 7 different software packages. These software packages were used to automatically segment the 75 HCP subjects used to build the atlas. Those automatic segmentation were later semi-automatically corrected and fused (and also manually corrected when needed). The seven software packages used are the following:

1. **vol2Brain** (Manjón et al., 2022): This software is able to segment the brain into 135 regions. It is based on non-local multi-atlas label fusion and is available as an online tool at https://volbrain.net.

2. **hypothalamus_seg** (Billot et al, 2020): This method is able to segment the hypothalamus into different parts using a convolutional neural network. The code is available on GitHub (https://github.com/BBillot/hypothalamus_seg).

3. **BrainVISA** (Mangin et al., 2004): BrainVisa is a software for brain segmentation and sulcal morphometry and was used to segment the brain sulci. The software is available at https://brainvisa.info. This software had several updates in the last years (Perrot et al., 2011; Rivière et al., 2022).

4. **Freesurfer** (Fischl et al, 2000)**:** Freesurfer is a well-known software package to segment the brain and analyze cortical thickness. The software can be downloaded here: https://surfer.nmr.mgh.harvard.edu. In the construction of the atlas, we used several subregion segmentation tools integrated into Freesurfer 7.3. Specifically, we used the tools for Brainstem (Iglesias et al., 2015a) and amygdala/hippocampus (Iglesias et al., 2015b) segmentation.

5. **pBrain** (Manjón et al, 2020)**:** This software is able to segment the 3 deep grey matter structures related to parkinsonism (substantia nigra, red nucleus and subthalamic nucleus). It is based on non-local multi-atlas label fusion and is available as an online tool at https://volbrain.net.

6. **HIPS** (Romero et al, 2017)**:** This software is able to segment the hippocampus subfields with 2 different delineation protocols (with 3 and 5 labels each). It is also based on non-local multi-atlas label fusion and is available as an online tool at https://volbrain.net.

7. **CERES** (Romero et al., 2017): CERES is an online automated software to segment the cerebellum lobules. It represents the current state of the art on cerebellum segmentation and is available as an online tool at https://volbrain.net.

## 2.4. Protocol integration process

To fully label each subject, we ran the described software and adapted the results to incrementally label them. The first applied pipeline was the vol2Brain software. This software segments a T1w image into 135 cortical and subcortical labels (Manjón et al., 2022). Because this method works at 1 mm$^3$ resolution and not at 0.125 mm$^3$ resolution, we decomposed the high-resolution T1w volume of size 362x434x362 voxels into 8 volumes of 181x217x181 voxels using a stride decomposition (step=2 in each dimension) (Manjon et al. 2020a). The eight volumes were segmented using vol2Brain and the resulting segmentations were composed back to the high-resolution space by inverting the striding operation over the label maps. This resulted in a single volume of size 362x434x362 fully labelled with 135 labels (we used this approach because it gave much better results than labelling at 1 mm$^3$ and later interpolating to 0.125 mm$^3$ resolution).

**Tissue error correction**

After this automatic segmentation, some systematic errors were present. Mainly, dura and vessels were misclassified as grey matter and sulcal CSF was underestimated. As these errors were found at the tissue level, the structure segmentation was automatically relabeled from 135 labels to 7 tissues (CSF, cortical GM, cerebral WM, deep GM, cerebellum GM, cerebellum WM and brainstem). To enforce the regularity of the different tissues their probability maps were filtered using a non-local means filter using as reference the T1w and T2w intensities to estimate the voxel similarities (this largely improved CSF underestimation). To correct the GM misclassification, we trained a UNET-like deep convolutional network using 12 cases manually corrected (this manual correction was performed using ITK-SNAP (Yushkevich et al., 2006)). The trained network was used to correct the remaining 63 subjects. Finally, the 75 cases were visually checked and the remaining errors were manually corrected. Once the tissues were corrected, we used these maps to automatically correct the original structure segmentation. To do so, a spatial diffusion process was used. In this process, all those voxels that did not change their tissue class were preserved but the remaining ones were relabelled. To perform this relabelling we used the spatial and intensity proximity to assign the new labels. Specifically, the probability map of each structure was smoothed using a 3D Gaussian kernel (to compute a spatial a priori probability map) and this information jointly with the voxel intensity was used to assign the new label. Basically, we used an iterative Bayesian Maximum A Posteriori (MAP) approach where the *a priori* probability was in the form of a spatial probability map and the likelihood of the intensities was calculated as the value of a Gaussian distribution at a given intensity with respect to the mean of the neighbor structures normalized by the variance of the structure intensities. An example result can be seen in Figure 2. The responsible for supervising this process were JVM and PC.

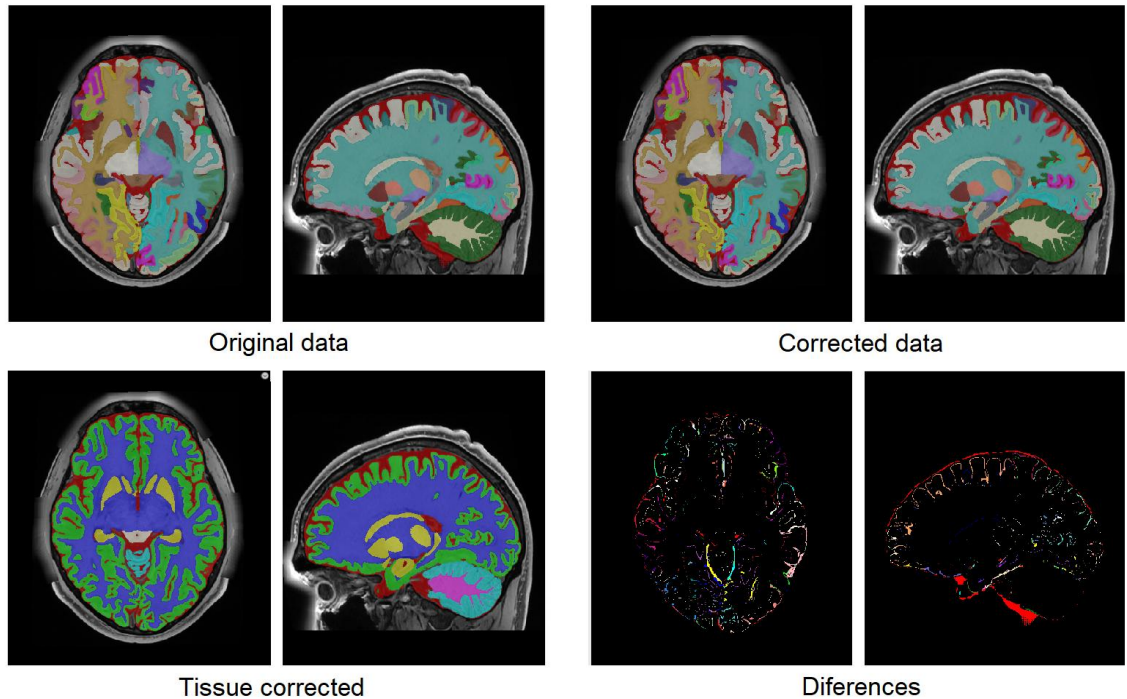

*Figure 2. Top-Left: Original segmentation before correction. Bottom-Left: Corrected tissue segmentation. Top-Right: Corrected segmentation using the corrected tissue maps. Bottom-Right: Difference between the original and corrected segmentation.*

**Hypothalamus integration**

The second protocol to integrate was the hypothalamus one. Hypothalamus_seg software segments the hypothalamus into 10 substructures. These labels were mapped on top of left and right white matter and partial volume voxels misclassified as WM in the border of hypothalamus/CSF were reclassified using the described spatial/intensity diffusion process. The added labels of the hypothalamus were: Right Anterior Inferior Hypothalamus, Left Anterior Inferior Hypothalamus, Right Anterior Superior Hypothalamus, Left Anterior Superior Hypothalamus, Right Posterior Hypothalamus, Left Posterior Hypothalamus, Right Tubular Inferior Hypothalamus, Left Tubular Inferior Hypothalamus, Right Tubular Superior Hypothalamus and Left Tubular Superior Hypothalamus. The responsible for supervising this process were JVM and PC.

**Sulci integration**

The cortical sulci were obtained from the T1w images via the following steps embedded into the Morphologist pipeline of BrainVISA [http://brainvisa.info, Mangin et al., 2004]. First, the brain mask was obtained with an automated skull stripping procedure including bias correction, histogram scale-space analysis and mathematical morphology. Second, the brain mask was split into hemispheres and segmented into grey matter, white matter and cerebrospinal fluid. Third, a negative cast of the cortical folds was segmented and labelled into sulci. The fold segmentation results from a 3D crevasse detector reconstructing each fold geometry as the medial surface from the two opposing gyral banks using a watershed procedure. A Bayesian pattern recognition approach relying on Statistical Probabilistic Anatomy Maps and multiscale spatial normalization was used to label the folds using a nomenclature of 124 sulci (Perrot et al., 2011).

To integrate this sulci information, the 3D external CSF volume (label 8) was relabelled into their closest sulci using automatic sulci segmentation (see Figure 3). After mapping the sulcal labels into label 8, the CSF not labelled in the proximity of sulcal labels was relabelled using the spatial-intensity diffusion process. The remaining external CSF was left unchanged (label 8) (mainly cerebellar and brainstem CSF and areas of the top of the brain far from the sulci). After the automatic segmentation, a visual QC was performed and small remaining errors were manually corrected using ITK-SNAP (mainly CSF voxels close to the cerebellum and brainstem). The responsible for supervising the whole sulci integration were EL, JFM and DR. In Figure 3 an example case of the labelling is shown.

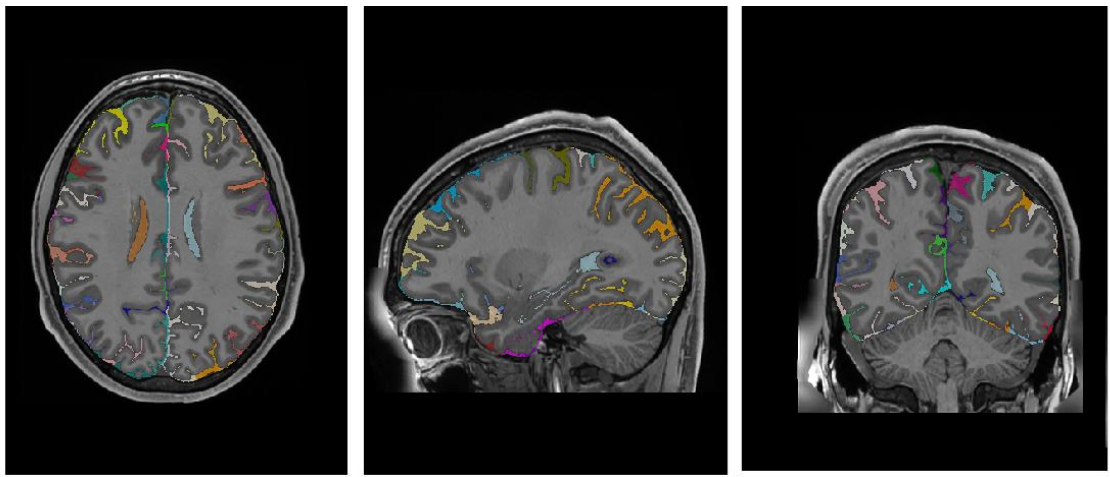

*Figure 3. Example of 3D sulci segmentation based on 2D mesh BrainVISA sulci labeling.*

**Brainstem and amygdala integration**

The next two structures to integrate were obtained using Freesurfer: brainstem and amygdala. Again, new labels for brainstem and amygdala were transferred to the new protocol and old labels for brainstem and amygdala were reclassified based on the bilateral spatial and intensity probabilities using the described MAP-based approach.

The Brainstem was divided into 4 regions: Midbrain, Pons, Medulla and Superior Cerebellar Peduncle. The new definition of the brainstem was slightly bigger than the vol2brain definition and overwrote some WM areas. Partial volume voxels were reclassified as brainstem or CSF again according to their intensity and location. Each amygdala was divided into 8 regions: Lateral Nucleus, Basal Nucleus, Central Nucleus, Medial Nucleus, Cortical Nucleus, Accessorial Nucleus, Cortico Amygdaline Transition and Paralaminar Nucleus. The new amygdala was also bigger than the vol2brain definition. Anterior Nucleus was not included in its definition as it was really small and occupied part of the entorhinal area. The amygdala segmentation had some systematic errors overestimating the frontal and inferior parts. In the Brainstem, the superior cerebellar peduncle was commonly overestimated and the medulla was underestimated in its inferior part. Manual corrections were performed when needed to correct these errors using ITK-SNAP. The responsible for supervising the brainstem and amygdala integration were MG and EL. Figure 4 shows an example of the relabelling process for the brainstem.

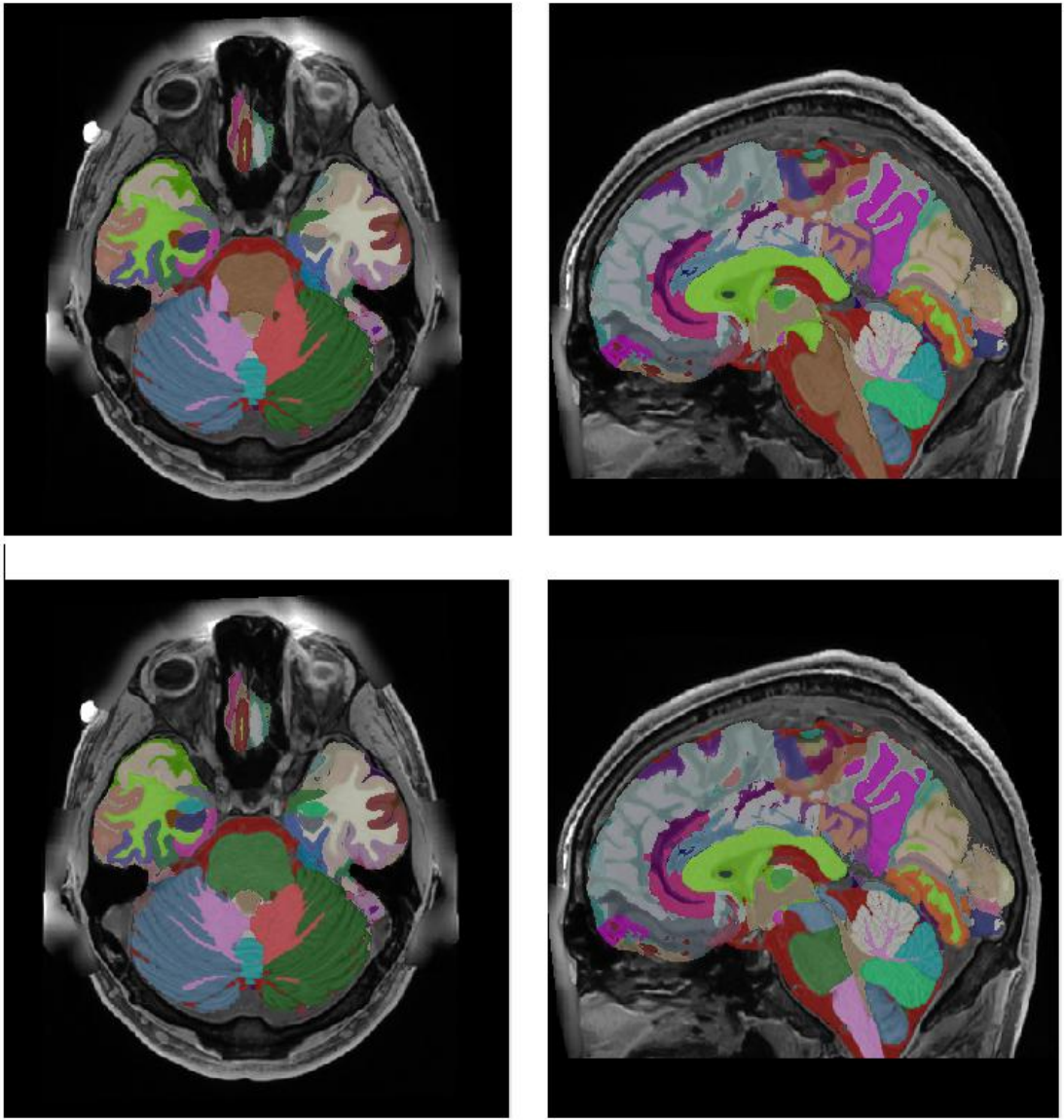

*Figure 4. Top row: original labels. Bottom row: integrated new labels. Note that the brainstem is now divided in its constitutive parts.*

**pBrain structures integration**

The pBrain pipeline provides segmentation of 3 structures of deep grey matter per hemisphere (red nucleus, substantia nigra and subthalamic nucleus) using a HR T2w image. As in the case of hypothalamus segmentation, labels were simply transferred over the corresponding white matter and midbrain region. The main problem of this integration was the fact that these structures partially occupied part of the midbrain (see Figure 5). Although these structures could be artificially divided to consider this fact we decided for simplicity to define a *partial* midbrain as a result of this integration process which was directed by JVM and EL.

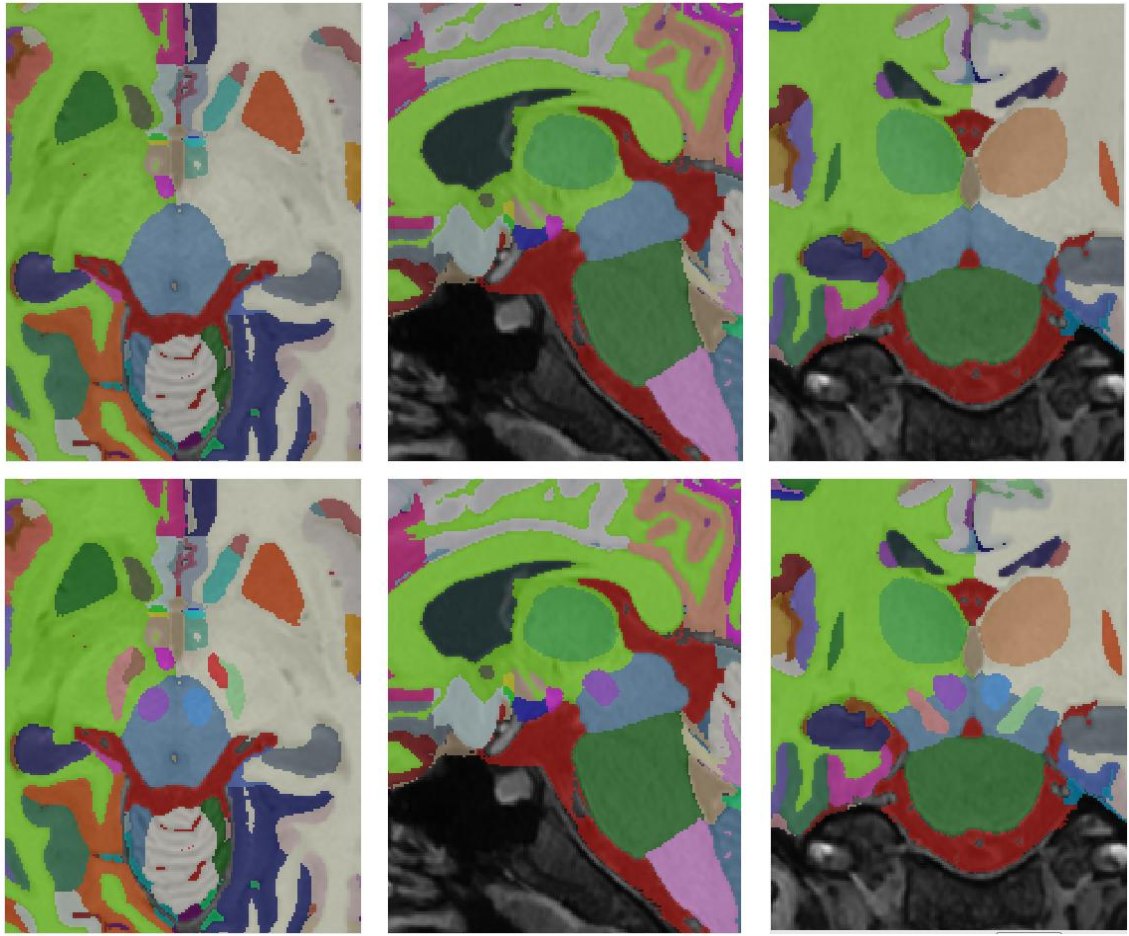

*Figure 5. Top row: original labels. Bottom row: integrated new subthalamic nuclei labels. Note that new labels modify the brainstem labels.*

**Hippocampus subfield integration**

To label the hippocampus subfields we used the software HIPS (Romero et al., 2017). HIPS segmented the hippocampus into 5 subfields (CA1, CA2/3, CA4/DG, SR/SL/SM and Subiculum) using the Winterburn protocol (Winterburn et al., 2013). However, we added two new structures, the Fimbria and the Hippocampal-Amygdalar Transition Area (HATA) to connect the hippocampus head with the amygdala. We obtained these labels from Freesurfer segmentation and fused with HIPS outputs. CSF pockets were classified as CSF. To map the new hippocampus definition over the vol2Brain-defined hippocampal area, first the vol2Brain hippocampus label was reclassified as WM and later the new definition was transferred only in the destination voxel has the WM label (to avoid amygdala label modifications). A throughout QC was done using ITK-SNAP and manual corrections were applied when necessary. The responsible to supervise the hippocampus integration were VP and EL. Figure 6 shows an example of the extended hippocampal protocol.

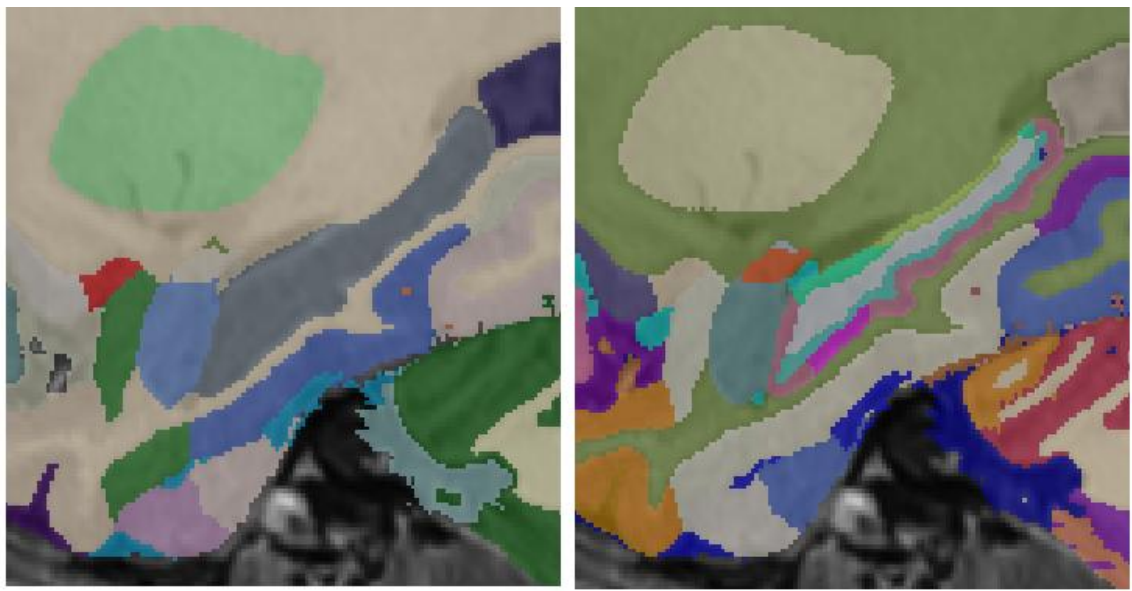

*Figure 6. Example of the old and new hippocampus subfield definition on top of a reference T1w MRI.*

**Thalamus subfield integration**

To obtain the thalamus nuclei we used the THOMAS software training dataset (https://github.com/thalamicseg/thomas_new). This dataset contains 20 White Matter Nulled (WMn) subjects. Specifically, we trained a deep neural network to segment WMn images and we used it to segment the 75 WMn images of our dataset. We used this network instead of using the THOMAS software because it provided better segmentations than THOMAS thus requiring less manual editing. Thomas protocol provides segmentation of 12 nuclei of the thalamus (Anterior Ventral Nucleus, Ventral Anterior Nucleus, Ventral Lateral Anterior Nucleus, Ventral Lateral Posterior Nucleus, Ventral Posterior Lateral Nucleus, Pulvinar Nucleus, Lateral Geniculate Nucleus, Medial Geniculate Nucleus, Centromedian Nucleus, Mediodorsal Nucleus, Habenular Nucleus and Mammillothalamic Tract). To label the entire thalamus as a whole we added a new label named *intermediate space,* defined as the WM region connecting all the nuclei, resulting in 13 total labels per thalamus (see Figure 7). In this case, the new thalamus definition was a bit smaller than the vol2Brain definition. The ones responsible for supervising the thalamus integration were MR, MG and TT. Again, the QC was performed using ITK-SNAP and manual correction was applied when needed (specifically, main errors were found at Mammillothalamic Tract and Habenular Nucleus which are the smaller nuclei).

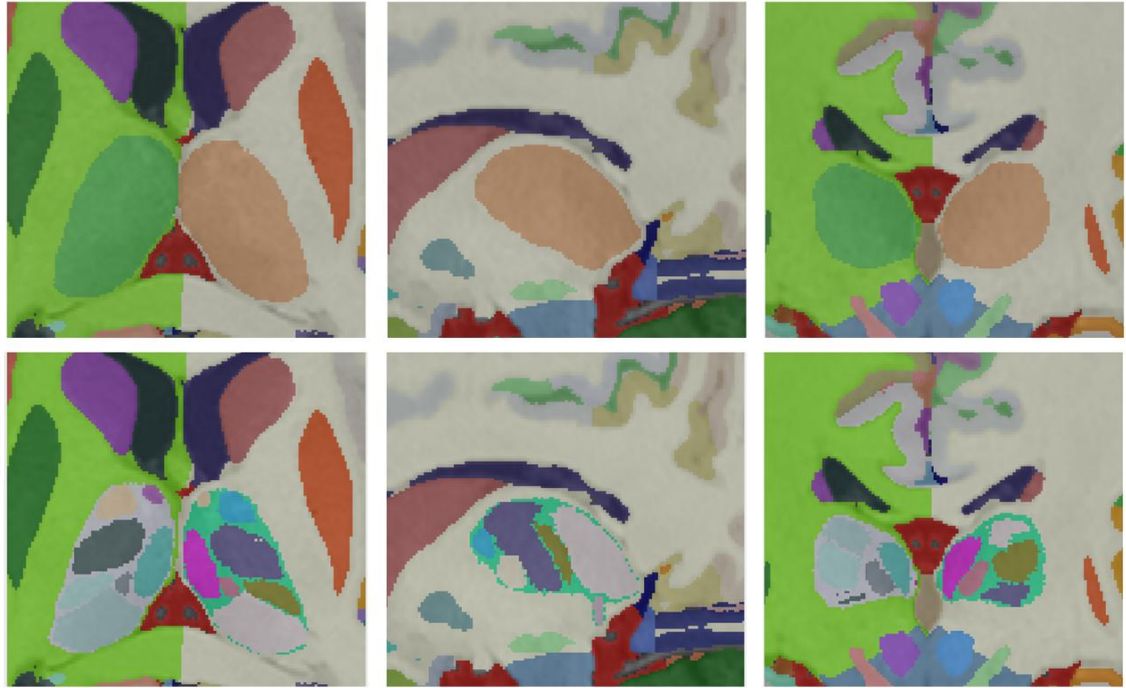

*Figure 7. Top row: original labels. Bottom row: integrated new thalamic nuclei labels.*

**Cerebellum lobules integration**

To segment the cerebellum, we used the CERES method (Romero et al., 2017). A multi-atlas path-based label fusion technique. CERES provides segmentation of 12 lobules of the cerebellum (Lobule I-II, Lobule III, Lobule IV Lobule V, Lobule VI, Lobule Crus I, Lobule Crus II, Lobule VIIB, Lobule VIIIA, Lobule VIIIB, Lobule IX and Lobule X) plus the cerebellar white matter (see figure 9). CERES works at 1 mm$^3$ resolution. Therefore, to generate the segmentation at 0.125 mm$^3$ resolution, we used the same striding-based approach we used previously. The new cerebellum definition is slightly smaller than the previous vol2Brain one, mainly due to WM label. We decided to increase the WM label combining both. The non-overlapping area was reclassified based on spatial and intensity information. Finally, the increased resolution of the images allowed to properly segment the intra-lobular white matter (not visible at 1 mm$^3$ resolution due to partial volume effects). This was done by SM as described in DeepCERES method (Morell-Ortega et al., 2025). Figure 8 shows an example of this integration process.

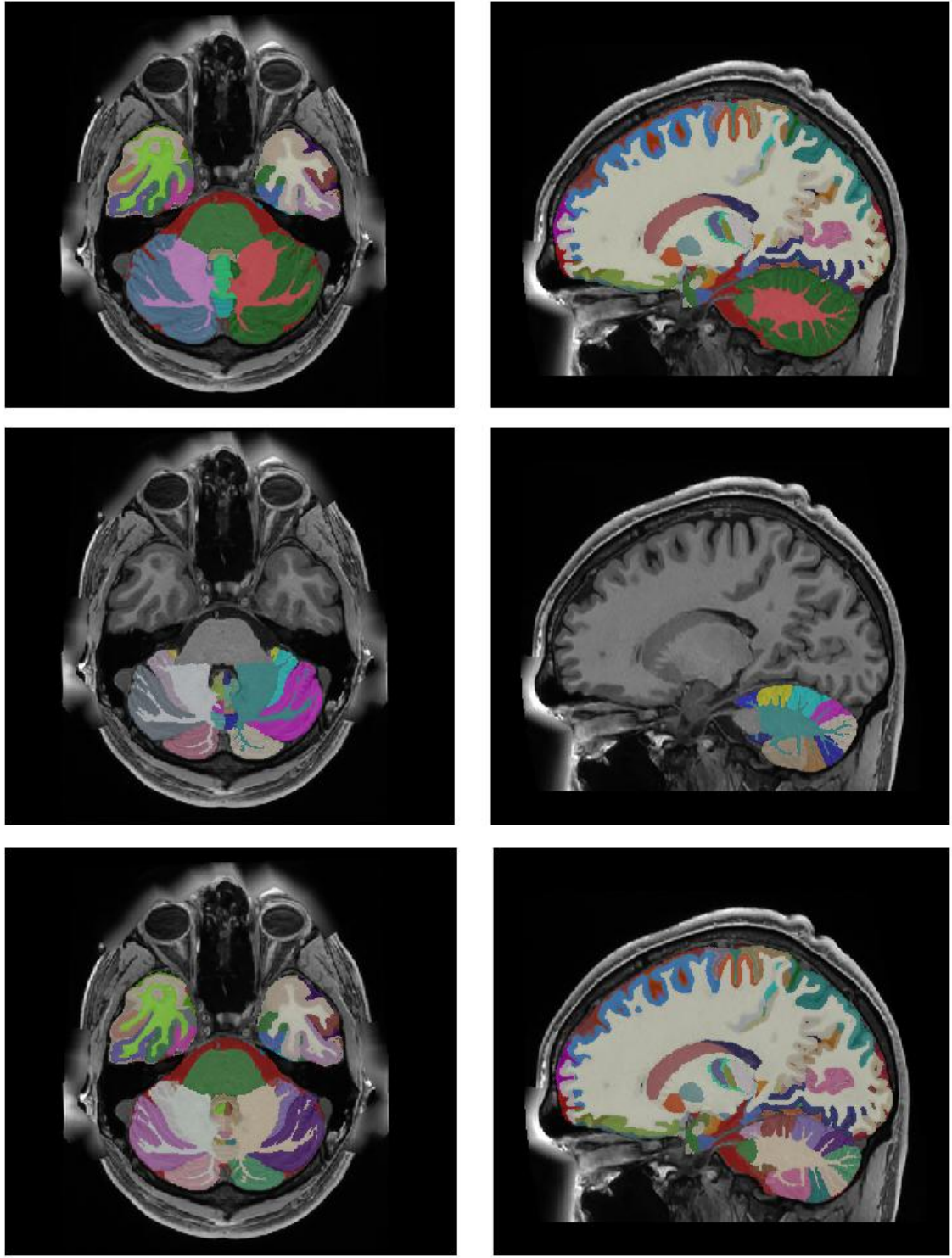

*Figure 8. Top row: original labels. Middle row: CERES segmentation. Bottom row: integrated new cerebellum labels.*

**Pallidum integration**

Finally, we decided to manually segment one structure for deep grey matter completeness. We divided the globus pallidum into its internal and external parts manually. This was done using ITK-SNAP from the entire pallidum segmentation obtained with vol2Brain pipeline. Figure 9 shows and example of the pallidum parcellation. This was done by JVM under the supervision of FA and MG.

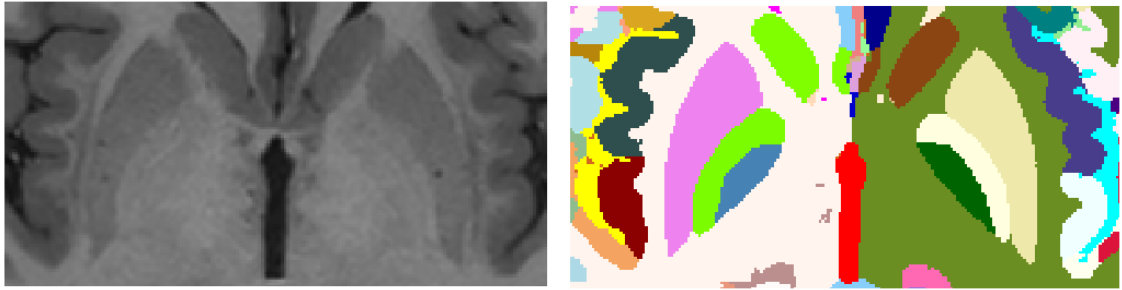

*Figure 9. Left: T1w image. Right: Pallidum segmentation of their internal and external parts.*

**Vessels and connective tissue**

After the whole integration process, we automatically performed a 3D hole-filling operation over the binary mask formed by the sum of all the defined structures (i.e. the foreground) to get a solid intracranial cavity volume (no holes). We named the label corresponding to the filled regions as "vessels+ connective tissue" as this label is mainly related to these tissues. Figure 10 shows an example of this label. This label allows to have a more compact brain anatomy and to fully define all the structures with the intracranial volume (ICV).

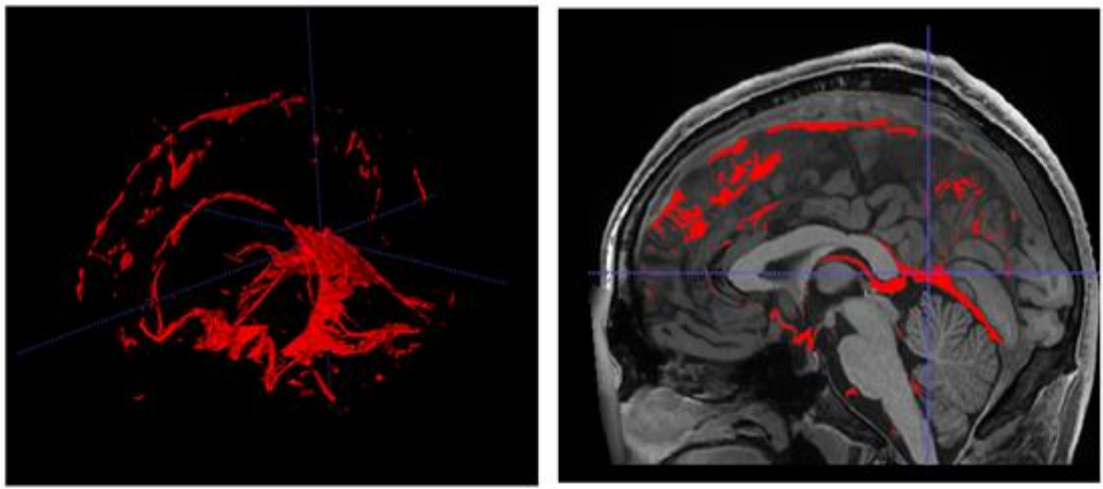

*Figure 10. Right: Example of Vessels and connective tissue. Left: 3D representation of this label.*

**Multiscale label generation**

After the whole integration process, we generate the corresponding label maps at multiple scales combining the corresponding labels (from substructure to structure, then to tissue and finally to organ (ICV). Figure 11 shows an example of the new holiBrain protocol. The substructure scale has 350 labels, the structure scale has 54 labels, the tissue scale has 9 labels and finally ICV has 1 label.

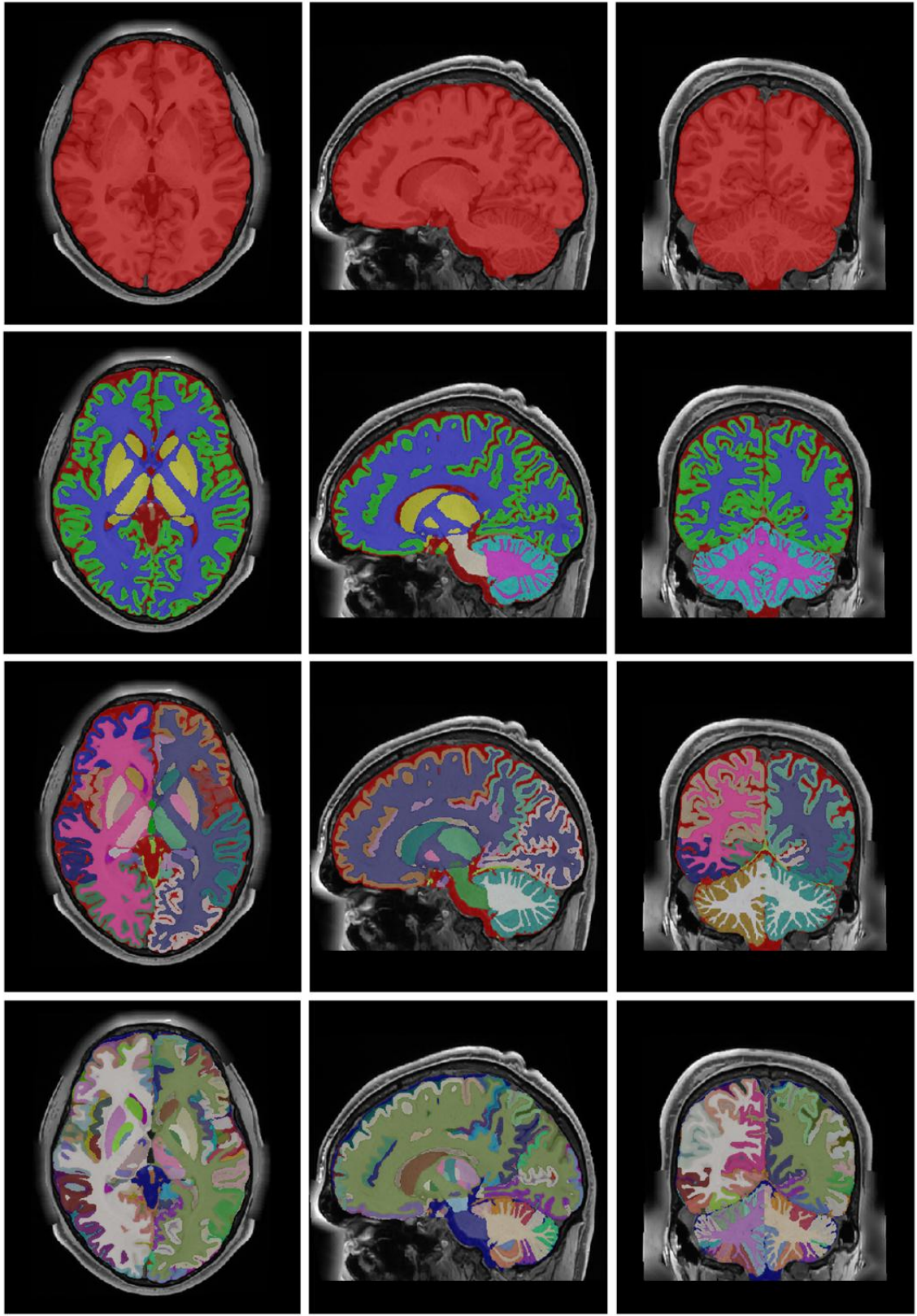

*Figure 11. From Top to bottom: ICV (1 label), tissues (9 labels), structures (54 labels) and substructures (350 labels).*

## 2.5. Atlas construction

To build the atlas, the 75 T1w, T2w and WMn MRI at 0.125 mm$^3$ resolution at MNI152 space and their corresponding labels were used. The initial reference image used for the template generation was the ICBM 2009b Nonlinear Asymmetric template interpolated at 0.125 mm$^3$ resolution. To create the T1w/T2w/WMn MRI templates, an iterative approach was employed. At each iteration, the T1w MRI underwent non-linear registration to the reference using the FireANTS software (Jena et al., 2025) to estimate the non-linear deformation field. To refine the reference image, the registered images were averaged, generating a new reference image at each step. Compared to ANTs (Avants et al., 2008) software, FireANTS provided an acceleration of 300x thanks to its GPU-based nature which resulted in several hours of processing compared to days with ANTS.

A total of 10 Iterations were performed for the template generation process. The template generation scripts used in this process are publicly available in the FireANTS repository (https://github.com/rohitrango/FireANTs). The deformation fields obtained from the T1w MRI template construction were subsequently employed for the generation of the average T2w and WMn MRI templates, as well as for the creation of the atlas labels. T1, T2 and WMn atlases were obtained using the median of the Laplacian-based sharpened images which reduces blurring of the atlases. A majority voting scheme was applied to all the registered label images to estimate the atlas labels at the different scales. Figure 12 shows a comparison of the proposed T1 atlas and the upsampled MNI152 T1 atlas to highlight the improved resolution of the proposed atlas.

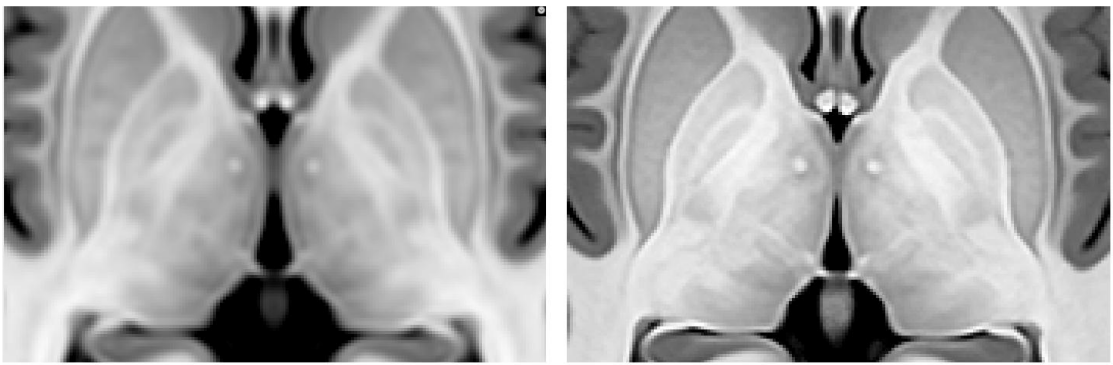

*Figure 12. Left: Central area of the upsampled MNI152 atlas. Right: Central area of the proposed atlas. Note that the proposed atlas has a better image quality compared to classic MNI152 atlas (for example, thalamic nuclei can be better observed at the new atlas).*

# 3. Results

After the whole integration process, we generate the last version at multiple scales (substructure/structure/tissue/organ). Figure 13 shows the holiAtlas templates for T1w, T2w and WMn MRI and the corresponding labels. Label definitions and relations among scales can be inspected in the appendix section. The generated holistic atlas jointly with the multiscale label definitions is publicly available through the following links: https://volbrain.net/public/data/holiatlas_v1.0.zip and https://zenodo.org/records/15690524 under a Creative Commons license.

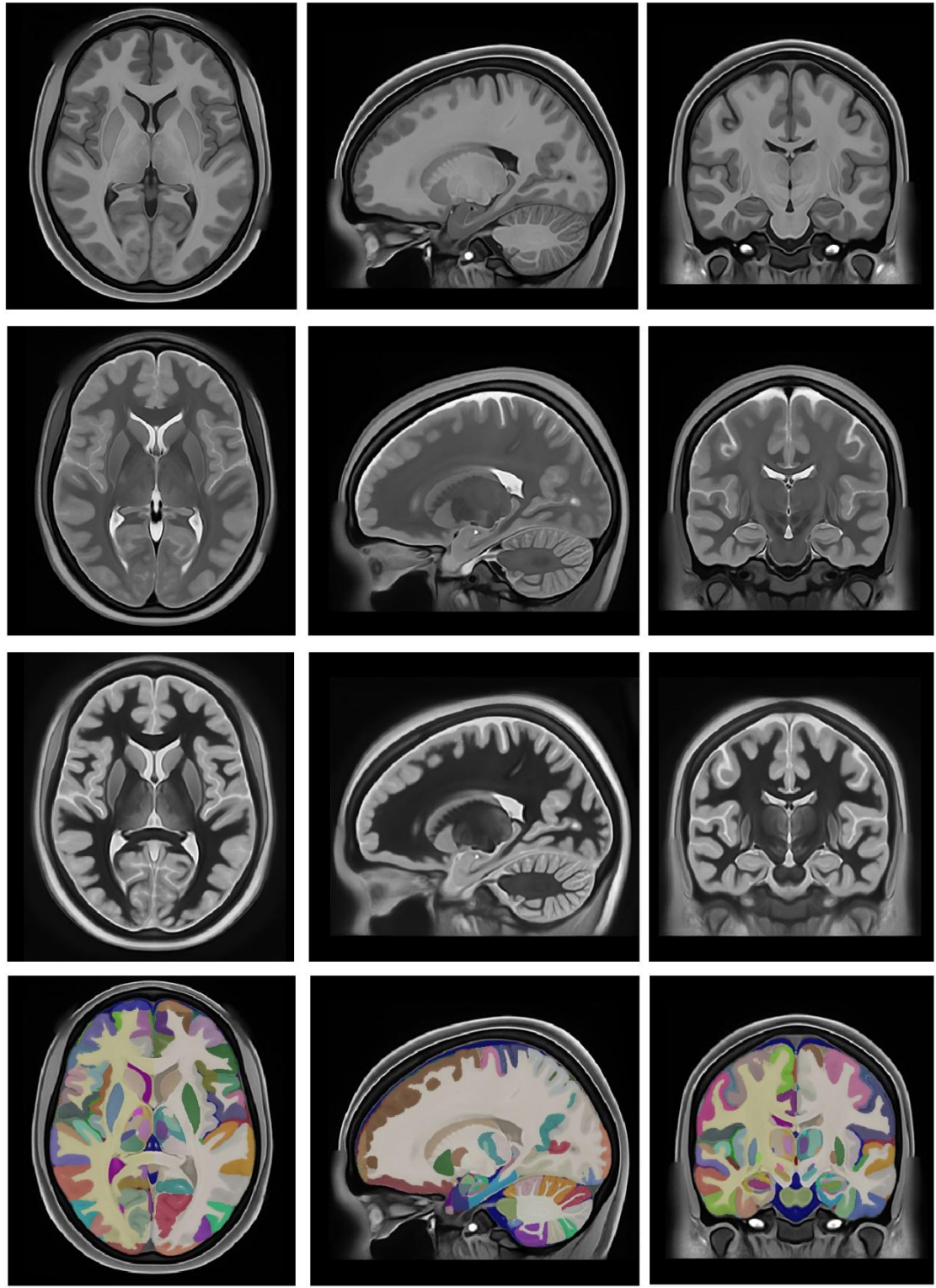

*Figure 13. From Top to bottom: Average T1w MRI template, average T2w MRI template, average WMn MRI template and majority voting atlas labels at substructure resolution.*

## 4. Discussion

This paper introduces holiAtlas, a comprehensive, multimodal, and ultra-high-resolution MRI-based atlas of the human brain anatomy. This atlas was constructed by fusing data from various local protocols with corresponding scales, resulting in a densely labelled protocol. The creation involved the use of images and segmentations from 75 healthy subjects, employing T1w, T2w, and WMn MRI contrasts at a 0.125 mm³ resolution. This atlas provides a holistic view of brain anatomy at multiple levels and scales, serving as a valuable resource for segmentation methods, research, and education.

The proposed atlas surpasses, most of the existing MRI-based brain atlases in resolution, providing detailed anatomical information. The conventional atlases, like the MNI152 Atlas, offer resolutions up to 1 mm³ (Mazziota et al., 2001), while holiAtlas achieves a significant leap to 0.125 mm³. This finer resolution enhances the precision of structural mapping and allows for a more specific exploration of brain architecture.

holiAtlas's unique contribution lies in the integration of diverse delineation protocols from 7 software packages. This integration process includes correction steps, such as tissue error correction, to refine and enhance the accuracy of the labels. This approach not only synthesizes information from different sources but also rectifies systematic errors, providing a more reliable atlas.

The multimodal nature of the atlas, encompassing T1w, T2w, and WMn MRI contrasts, broadens its descriptive power as it offers a multiple view of the same organ. The holiAtlas's multiscale label generation facilitates versatile usage, accommodating multiple research needs, from substructure to overall organ analysis.

Incorporating synthetic image synthesis techniques, such as generating WMn-like images using data from HCP1200, showcases the paper's innovative approach to overcoming data limitations. This technique expands the usability of the atlas by providing a complete set of contrasts even when not all contrasts are available in the original datasets. In the future, we will add other synthetic modalities such as FLAIR or contrast-enhanced T1.

During the creation of the atlas, which lasted nearly 3 years and employed the expertise of many people, we emphasized the importance of manual correction and quality control in the integration process. While automated methods are powerful, the inclusion of human expertise ensures the accuracy and reliability of the final atlas. The main issue was to fuse partial (e.g., focused on specific areas or defined at a specific scale) and incompatible (e.g., a given voxel can be assigned to different structures according to the used methods) protocols into a consistent and coherent protocol across scales to obtain the final holiBrain protocol. This iterative correction process enhances the atlas validity for a broad range of applications.

Differently from other atlases, the proposed atlas is based on what you can actually see and measure in real in-vivo MR images instead of using histological images that although having an impressive resolution do not provide a direct match to the acquired MR images.

We believe that the increased resolution and label density of the proposed atlas may help improving the early detection of neurological diseases by focusing on substructure atrophies rather than structure atrophies which are more difficult to detect. For instance, the characteristic atrophy of the amygdala and the hippocampus in Alzheimer's disease (Coupé et al., 2019; Planche et al., 2022) is due mainly to the volume reduction of the lateral and cortical amygdaloid nuclei (González- Rodríguez et al., 2023), and to neurodegeneration in the CA1 subfield (West et al., 2000; Kril et al., 2004), and therefore the analysis at the substructure level is necessary. A similar situation is observed in Parkinson's disease, where a volume reduction is present in the hippocampus and putamen (Tanner et al., 2017) or in the different clinical variants of frontotemporal dementia (Planche et al., 2023b). In summary, the high resolution and increased label density of the present atlas will facilitate the analysis of particular cases at the substructure level.

The proposed atlas may play a central role in structural MRI analysis, serving as a foundational resource across a wide range of neuroimaging applications. One key use case is automated brain structure segmentation, where the atlas provides anatomical priors that guide the classification of grey matter, white matter, and cerebrospinal fluid, improving segmentation accuracy, particularly in challenging datasets. Atlases are also essential in disease mapping, acting as common coordinate systems that enable voxel-wise comparisons between populations, thereby facilitating the identification of disease-specific patterns in disorders such as Alzheimer's disease, multiple sclerosis, or brain tumors. Moreover, longitudinal studies rely on atlases to provide consistent spatial references over time, allowing accurate tracking of brain changes due to development, aging, or neurodegeneration. In the context of machine learning, atlases serve as powerful tools for weakly supervised learning, self-supervised representation learning, or template-based data augmentation. Finally, they support the creation of population-level brain maps, helping researchers model structural variability across demographics and build normative anatomical models (similar to AAL atlas).

We acknowledge the limitations of the proposed atlas. The fact that it is based on the fusion of automatic segmentations may raise doubts about its accuracy (despite our exhaustive QC process). However, manual segmentations also have their limitations. Systematic and random errors can be present in each case. However, modern medical image analysis software has a variability close to the inter-rater variability which somehow justifies the use of this approach (Romero et al., 2017; Coupe et al., 2020). Even in the case of systematic errors, if these errors systematically over or underestimate the volume of interest for healthy and diseased brains in the same manner the effect size induced by the disease can be effectively captured (Tustison et al., 2014). Besides, the data comes from a young cohort and therefore may not be representative of pediatric or elderly populations. Finally, the ultra-high resolution of this atlas may not drastically improve the analysis of large structures, we think that it can definitively help the study of small structures (e.g., nuclei or lobules).

## 5. Conclusion

In conclusion, the holiAtlas presented in this paper stands as a novel contribution to the field of neuroscientific research and brain imaging. The integration of multiple protocols, ultra-high resolution, and the synthesis of missing modalities make it a unique and valuable resource. The atlas construction methodology, combining automated segmentation with meticulous manual correction, ensures a high level of accuracy. We have addressed many limitations of existing atlases, providing a comprehensive tool that can drive advancements in segmentation methods, facilitate diverse research endeavors, and serve as an educational reference. The multimodal, multiscale, and ultra-high-resolution features of holiAtlas position it as a significant asset for the neuroscience community, offering new avenues for understanding the intricate complexities of the human brain.

## Acknowledgments

This work has been developed thanks to the projects PID2020-118608RB-I00 and PID2023-152127OB-I00 of the Ministerio de Ciencia e Innovacion de España. This work also benefited from the support of the projects DeepvolBrain, HoliBrain and FOLDDICO of the French National Research Agency (ANR-18-CE45-0013, ANR-23-CE45-0020-01 and ANR-20-CHIA-0027-01). Finally, this study received financial support from the French government in the framework of the University of Bordeaux's France 2030 program / RRI "IMPACT, the PEPR StratifyAging and the IHU VBHI (ANR-23-IAHU-0001).